# Probabilistic sequence alignments: realistic models with efficient algorithms

Edouard Yeramian* and Edouard Debonneuil

*Corresponding author, mail: yeramian@pasteur.fr

Unité de Bioinformatique Structurale, CNRS URA 2185
Institut Pasteur, 25-28 rue du Dr Roux, F-75724 Paris CEDEX 15, France

PACS numbers: 87.10.+e, 87.15.Cc

Alignment algorithms usually rely on simplified models of gaps for computational efficiency. Based on an isomorphism between alignments and physical helix-coil models, we show in statistical mechanics that alignments with realistic laws for gaps can be computed with fast algorithms. Improved performances of probabilistic alignments with realistic models of gaps are illustrated. Probabilistic and optimization formulations are compared, with potential implications in many fields and perspectives for computationally efficient extensions to Markov models with realistic long-range interactions.

Even though the alignment of sequences is a routine task in biology many practitioners may not question closely how comparisons are made, and current methods suffer from several drawbacks. One limitation concerns the use of simplified models for gaps in sequences to render calculations tractable. Sequence alignment programs implement either heuristic or exhaustive schemes, such as in the Needelman-Wunch method (see for example [1]). Here we examine the exhaustive schemes, which are classically formulated as 'dynamic programming' algorithms. They consist either of optimization schemes which find the 'best' alignment for a given model, or of probabilistic schemes based on partition functions -as pioneered by Miyazawa [2]- in which all alignments, with their respective weights, are evaluated. In practice optimization schemes are largely predominant today in various bioinformatics fields. Apart from questions on parameter refinement, there are four critical issues for improving sequence alignments: 1) Non-affine gap scores. 2) Allowance for gap overlaps. 3) Obtaining alternative solutions, beyond those with optimal scores. 4) Tractable and efficient algorithms. Here we show, in a probabilistic context, that all four requirements may be satisfied, by resorting to a strict correspondence between bioinformatics models for sequence alignments [1] and biophysics models for DNA helix-coil transitions [3]. In addition, we find that the effect of realistic non-linear gaps is not equivalent in the optimization and probabilistic contexts. We show with examples that the corresponding performances are significantly poorer in the optimization context. Based on these conclusions, we examine perspectives for efficient probabilistic sequence alignments using realistic gap laws. Finally we suggest that these ideas might be extended to Markovian modellings (HMM, etc), in numerous fields, with quite general long-range interactions.

*Backround for sequence alignments* - For gap penalizations, the need for non-affine laws was demonstrated in studies of DNA based on the comparison of human and mouse genomes [4], and of proteins based on structural studies, with a power-law fit in [5] and a multiexponential fit in [6]. Today most optimization-based sequence alignment methods resort to affine gaps, except for a few methods specifically designed to handle non-affine monotonic gaps with efficient algorithms (for example [7]). In the probabilistic frame general gaps were introduced [8] at the price of increased, unaffordable, complexities. For realistic modelling of sequence evolutions, gap overlaps must also be considered [4]. However, for algorithmic constraints, such overlaps are forbidden in most alignment methods. For affine gaps only, in the optimization context, the "generalized affine gap model" [9] was developed to handle overlaps. In the alignment problem, as in problems such as RNA secondary structures, the realisation that biologically relevant solutions may not correspond to optimal solutions, led to the development of optimization methods yielding sub-optimal solutions (for example [1]). Finally, efficient algorithms are required for biologically significant sequence lengths but this requirement usually conflicts with the need to represent long-range effects realistically (length-dependent gap penalizations). This conflict has typically been resolved by various simplifications.

*Correspondence between alignment and helix-coil models* - In order to fulfill, in the probabilitic context, the four requirements above for efficient and realistic alignments we first develop formal correspondences between the two alignment and helix-coil models. We note that the requirement 3) above for alternative solutions is naturally satisfied in the probabilistic context, with the obtention of all solutions, with no discrimination of sub-optimal classes.

On formal grounds, certain correspondences between helix-coil and alignment models have already been examined (for example [8]). Fig. 1 illustrates more detailed correspondences, which underly the algorithmic ideas we will develop. The basic isomorphism is between linear helix-coil model and basic sequence alignment (Fig.1a and 1b, respectively). Base pair stackings correspond to substitution weights and denaturation loops to gaps. In the helix-coil model polymer physics dictates power-law representations for loop entropies, is strict correspondence with the empirically derived non-linear laws for gaps. In the basic alignment gap overlaps are forbidden, with 'forced' homologies between gaps in the two sequences (Fig. 1b). In this simple correspondence the calculation burden derives from the need for combinatorial enumerations of loop (respectively gap) lengths (L1, Fig. 1) for proper weight attributions. When gap overlaps are included, the calculation increases by one order of complexity due to the need for a double combinatorics: for any length L1 of the first combinatorics, we mut consider the full combinatorial enumeration of lengths L2 (for proper weight attributions, depending on both lengths L1 and L2; in green and magenta respectively, Fig. 1). The methodological problem in handling such double combinatorics is equivalent to the one encountered in the biophysical model, with the introduction of circularity constraint in the linear helix-coil: free-ends (or 'dangles') must be joined for loops of lengths L2+L2' (Fig. 1c). The same type of constraints is also encountered in linear helix-coil models which permit non-symetrical loops (Fig. 1d).

*Complexities and algorithms* - For complexities associated with combinatorial enumerations shown in Fig. 1, adoption of affine laws for gaps in optimization dynamic programming is necessary due to the additivity property of calculations with the Max operator. In the dual multiplicative frame (usually probabilistic, but also possibly for optimization formulations), through exponentiation of various terms (scores etc) in the calculations, affine penalizations for gaps are replaced by exponential weights, and additivity property tranforms into a multiplicative one, following the fundamental property of the exponential function (exp(x)exp(y)=exp(x+y)). This property permits propagation of the calculations in the multiplicative frame following a nearest-neighbour scheme, with reduced complexity, even though the penalization is length-dependent, and so of long-range type. Derivations here depend on the extension of this idea to long-range effects, or gaps, which in principle could be of any kind. The rationale for this extension is that the long-range effect can be represented numerically as a sum of N exponentials following principles in the SIMEX method (SIMulations with EXponentials [10,11]). Partition function calculations can then be split into N terms, with the benefit of the basic exponential propagation in the calculation of each term. In general, appropriate multiexponential representations can be obtained using the Padé-Laplace method in signal analysis [12]. For the alignment problem such representation is available, on empirical bases ([6]; with N=4). The possibility to reduce calculations by one order of complexity with multiexponential representations of long-range effects was first formulated in the Fixman-Freire (FF [13]) method, for the linear helix-coil model. Based on this idea, the SIMEX method was formulated in two steps: 1) reformulation of the FF in terms of partition function calculations [10], rather than specific conditional probabilities, for models with simple combinatorics, and 2) extension to higher-order models with multiple combinatorics [11] such as circular helix-coils (Fig. 1c). The extended formulation [11] was transposed to the model shown in Fig. 1d [14]. Use of the SIMEX formulation, permits reduction of algorithmic complexities by one, or two or more, orders of complexities for models involving single, or two or more, mutually dependent combinatorics. The method is limited to partition function calculations with long-range effects that depend only on number of elements in a given state. Elements do not need to be contiguous, as in gaps or loops. The method can be applied when long-range effects can be expressed numerically as sums of a limited number of exponentials. Then, due to the exponential propagation property, the solution for reduced complexity is conceptually unique. We note that concepts underlying this solution cannot be extended to the optimization framework.

The fundamental point here is that the formal isomorphisms we have described, permit exploitation for the alignment problem of algorithmic concepts elaborated for biophysics problems with a reduction in algorithmic complexities as in [10] and [11]. We consider two sequences, of lengths n and m respectively, and we want to evaluate probabilities P(i,j) that elements (amino acids or bases) at position i in the first sequence and position j in the second sequence are aligned. Probabilities are calculated as the ratios P(i,j)=Z(i,j)/Z, with Z the total partition function for all alignments, and Z(i,j) the partition function restricted to configurations with i and j aligned. We adopt the same configurational representations as for DNA physics models (Fig. 2; [10,11]): symbol '1' is for two elements in the aligned state; symbol '0' for the non-aligned state (a gap corresponding to a stretch of 0's, limited on both sides by 1s; stretches of 0's at the ends correspond to 'dangles' which can be penalized according to the models), and finally the symbol 'X' represents a 'wildcard', corresponding to either state '0' or '1'. In the optimization context, penalized dangles are associated with 'global' alignments, whereas in local alignments dangles are not penalized, as in linear helix-coil model with 'free-ends'. If $Z_f(i,j)$ is the partition function associated with alignments up to position i on the first sequence and j on the second sequence, with i and j aligned (Fig. 2a), and calculations are performed processively, from left to right ('forward'):

$$Z_f(i,j) = \left[ Z_f(i-1,j-1) + \sum_{k=1}^{i-2} Z_f(i-k,j-1)\omega(k) + \sum_{l=1}^{j-2} Z_f(i-1,j-l)\omega(l) + \sum_{k=1}^{i-2}\sum_{l=1}^{j-2} Z_f(i-k,j-l)\omega(k,l) + 1 \right] s_{i,j}$$

with $s_{ij}$ the substitution weight for i and j aligned. For the general model, with gap overlaps, the calculation of the partition function, as the sum of the nm $Z_f(i,j)$ terms, requires $n^2m^2$ operations (with nm operations for the calculation of each $Z_f(i,j)$). It is easy to see that in a strictly processive treatment, for example from left to right, the calculation of the full set of probabilities P(i,j) is in complexity $n^3m^3$. Algorithmic complexities are then reduced in two ways. The first conforms to a forward-backward formulation (Fig. 2b), as used in classical dynamic programming implementations (such as HMM; see for example [1]), with the calculations proceeding from both sides, and equations for backward calculations ($Z_b(i,j)$ terms) rigorously similar to the one above. Then, $Z(i,j) = Z_f(i,j) Z_b(i,j)$, and the complexity for the evaluation of probabilities set is $n^2m^2$ (recursive steps in the one-way scheme are avoided). Secondly, complexity is reduced by applying SIMEX principles, with multiexponential representation of the gap function, and the corresponding reformulation of the recursive calculations for the benefit of numerical factorings following the fundamental exponential property. In the equation above, the application of simple and generalized SIMEX ideas [10,11], results in a reduction of calculations with the two single sums by one order of complexity, with the variable indexes k and l replaced by the index N (N=4; following the number of exponentials). Similarly, the calculation with the double sum is reduced by two orders of complexity, with the double indexing replaced by a mere NxN index. Using the forward-backward treatment and applying the SIMEX principle for both forward and backward equations, the calculation of the full set of probabilities for the model with gap overlaps is in complexity $N^2nm$, similar to calculations with only nearest-neighbour interactions. Computation times for long sequences are dramatically reduced.

*Examples and conclusions* – For applications, we highlight major conclusions with examples. We neglect gap overlaps for the moment. (Treatment of this point, along with detailed benchmarks will be presented elsewhere). Our conclusions are illustrated with examples in Fig. 3 and can be summarized as: 1) non-linear gaps can be indispensable for correct probabilistic alignments (for increasingly more difficult cases, notably with large gaps); 2) non-linear gaps appear to be much less effective in the optimization context than in the probabilistic one; and 3) in both optimization and probabilistic contexts, the penalization of dangles can be critical with linear gaps and much less so with non-linear ones.

The need for non-linear gaps, in probabilistic alignments, is shown in Fig. 3 with proteic sequences PTGA and psta: in Fig. 3e (close-up view in 3f) the complete structural alignment (see [15]) is obtained with non-linear gaps (the 4 exponential terms following [6]; the result being independant of the penalization of dangles) whereas only partial alignments are obtained with linear gaps (single exponential in the probabilistic context) in Fig. 3c (dangles non penalized) and in Fig. 3d (dangles penalized). In the optimization context, the alignment obtained with linear gaps (Fig. 3a, dangles penalized; without penalizations the alignment is completely disrupted) is comparable to that obtained in the probabilistic context with the same linear law (Fig. 3d), whereas with non-linear gaps the alignment (Fig. 3b; same alignment with dangles penalized) is much less satisfactory than that obtained in the probabilistic context (Fig. 3e). We note that, in the optimization context, non-linear, as opposed to linear, gaps tend to disrupt rather than imrove the alignments. Disruption still occurs with non-linear gaps in the optimization context, when linear gaps suffice to obtain correct structural alignments, in both probabilistic or optimization contexts. This is evident in comparing the linear optimization alignment of 1a12A and 1jtdB in Fig. 3g (similar probabilistic result) and the non-linear one in Fig. 3h, where the non-linear probabilistic alignment is strictly equivalent to the linear one.

*Perspectives* - The conceptual advance here, permitting the criteria for realistic sequence alignments to be satisfied simultaneously, is based on the establishment of solid links between structural biophysics and bioinformatics. Similar bridges between these fields have provided comparable insights in the ab initio identification of genes on structural physical bases [17]. The formalism developed here may also be fruitful in fields where Markovian models (such as HMM) resort to simplified representations for long-range effects (single exponential), for the same complexity reasons as for this problem. A combination of the SIMEX [10,11] and Padé-Laplace [12] methods should permit the introduction of realistic long-range effects in such models, without greatly increasing calculation times. Finally, the very different performances obtained with non-linear realistic laws in probabilistic and optimization contexts may correspond to rather general properties. It will be interesting to explore these differences in fields where such dual representations are important.

**Acknowledgments:** we thank Michael Nilges for constant scientific support and helpful discussions, Richard Miles and Olivier Martin for careful reading of the manuscript. The work was supported by the Ministèrere de la Recherche Scientifique (ACI-IMPBIO-2004– 98-GENEPHYS) and the Institut Pasteur (DVPI contract and Strategic Horizontal Programme on Anopheles gambiae).

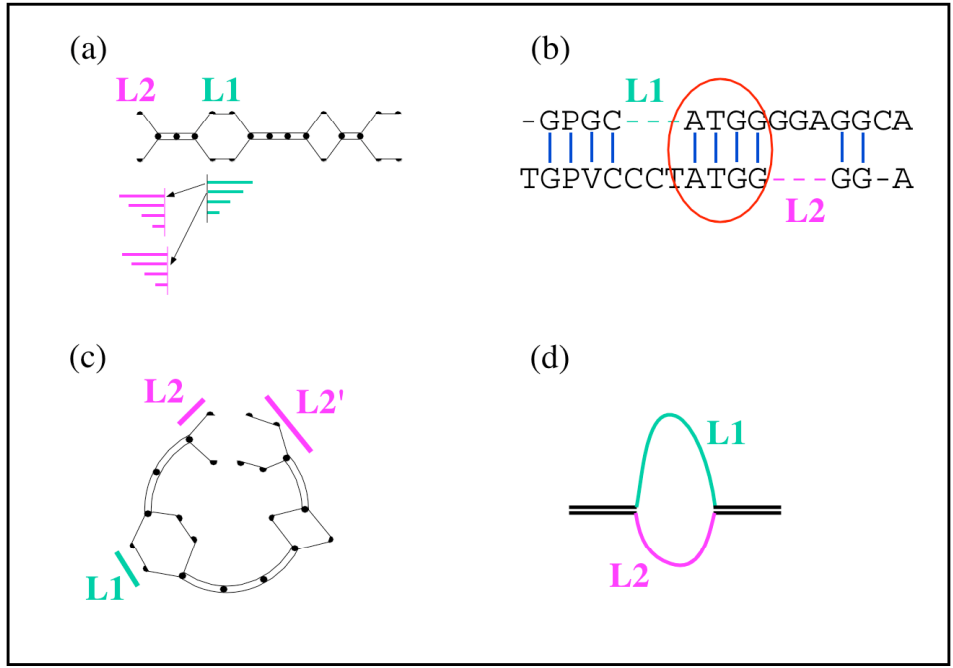

**Fig. 1.** Correspondences between biophysics (helix-coil) and bioinformatics (alignment) models. (a) Classical linear helix-coil model. Implementation of the model involves the enumeration (in configurational descriptions) of possible loop lengths (L1), for proper weight attributions. (b) Alignment model, with gaps (lengths L1 and L2) instead of loops. In simple models overlaps between gaps are forbidden, with imposed homology regions (such as the one circled in red). (c) Circular helix-coil model, involving a double combinatorics of length-dependent enumerations (for each length L1, the full enumeration of L2 lengths), for proper loop closures (lengths L2+L2') in circular configurations. (d) Linear helix-coil model with allowance for non-symetrical loops.

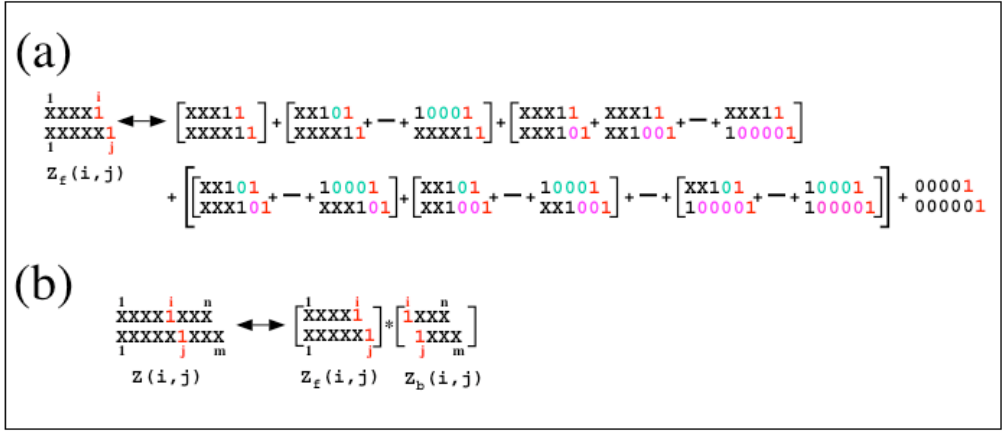

**Fig. 2.** Configuration representation for the alignment partition function calculations. Symbol '1' corresponds to elements in the aligned state (corresponding positions i and j are not necessarily in correspondence because of different sequence lengths), '0' to non-aligned state and 'X' as 'wildcard' (both states 0 and 1). (a) Calculation of forward partition functions $Z_f(i,j)$. (b) Forward-backward scheme.

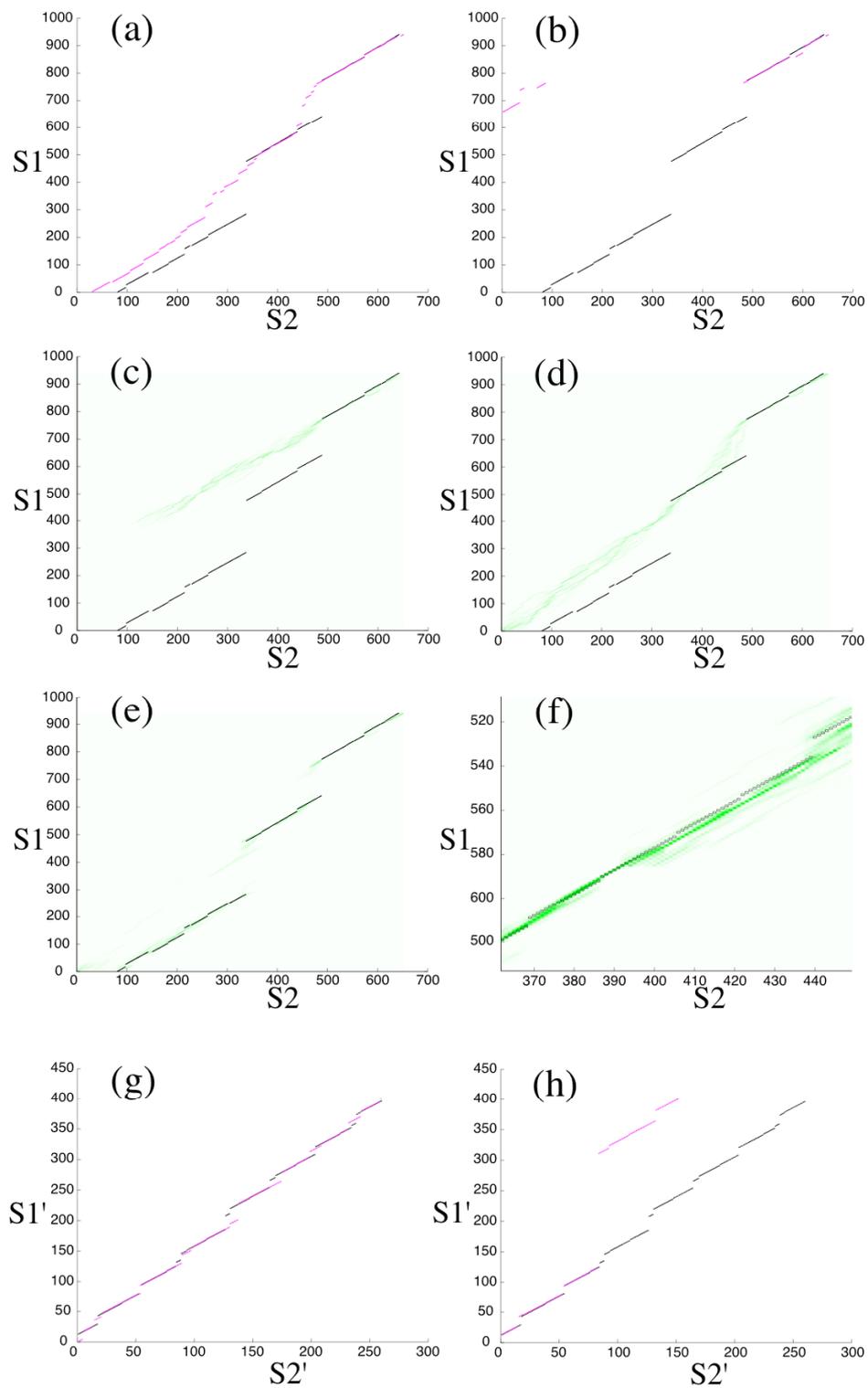

**Fig. 3.** Alignment dotplots for proteic sequences (PTGA (S1) and ptsa (S2); 1a12A (S1') and 1jtdB (S2')): reference, structural, alignments are in black, optimization in purple and probabilistic in green. For sequences in the range of 1000 amino acids, optimization calculations with non-linear gaps are performed without accelerations; for linear gaps the alignments are identical to those obtained with classical programs such as mglobalS [16]. (a) Optimization with linear gaps, dangles penalized. (b) Same as (a) but with nonlinear gaps. (c) Probabilistic with linear gaps, no penalization for dangles. (d) Same as (c) but with dangles penalized. (e) Probabilistic with nonlinear gaps, dangles penalized. (f) Close-up view of (e). (g) Optimization with linear gaps, dangles penalized. (h) Same as (g) but with nonlinear gaps.